\documentclass[10pt]{article}

\usepackage[paper=a4paper,dvips,top=1.5cm,left=1.5cm,right=1.5cm,
    foot=1cm,bottom=1.5cm]{geometry}

\usepackage{times}

\usepackage[fleqn]{amsmath}
\usepackage{amsfonts}
\usepackage{color}
\usepackage{amssymb}
\usepackage{amsthm}
\usepackage{amsopn}
\usepackage{xspace}
\usepackage{array}
\usepackage{epsfig}
\usepackage{bm}

\newcommand{\BE}{\begin{equation}}
\newcommand{\EE}{\end{equation}}
\definecolor{darkgreen}{rgb}{0.0,0.4,0.0}

\begin{document}

\begin{titlepage}

\vspace*{1mm}
\begin{center}

\vskip 1 pt


{\LARGE\bf Additional evidence of a new 690 GeV scalar resonance }

\end{center}

\begin{center}

\vspace*{14mm}{\Large M.~Consoli$^{(a)}$, L.~Cosmai$^{(b)}$, F.~Fabbri$^{(c)}$,
and G.~Rupp$^{(d)}$}
\vspace*{4mm}\\
{a) Istituto Nazionale di Fisica Nucleare, Sezione di Catania, Italy ~~~~~\\
 b) Istituto Nazionale di Fisica Nucleare, Sezione di Bari, Italy ~~~~~~~~~\\
 c) Istituto Nazionale di Fisica Nucleare, Sezione di Bologna, Italy ~~~\\
 d) CFTP, Instituto Superior T\'{e}cnico, Universidade de Lisboa, Lisboa,
Portugal}
\end{center}

\begin{center}
{\bf Abstract}
\end{center}
\par \noindent
An alternative to the idea of a metastable electroweak vacuum would be an
initial restriction to the pure scalar sector of the Standard Model,
but describing spontaneous symmetry breaking consistently with studies
indicating that there are two different mass scales in the problem: a mass
scale $M_H$ associated with the zero-point energy and a mass scale $m_h$
defined by the quadratic shape of the potential at its minimum. Therefore,
differently from perturbation theory where these two mass scales coincide,  
the Higgs field could exhibit a second resonance with mass
$(M_H)^{\rm Theor} = 690\,(30)$ GeV. This stabilises the potential, but 
the heavy Higgs $H$ would couple to longitudinal $W$s with the same typical
strength as the low-mass state with $m_h=125$~GeV and so would still remain
a relatively narrow resonance. 

While interesting signals from LHC experiments were previously
pointed out, we have now enlarged our data sample, sharpened the analysis of
some final states, and noted correlations between different channels that
point directly to such a second resonance. The combined statistical evidence,
even if roughly estimated, is thus so large that the observed deviations from
the background cannot represent statistical fluctuations.




\end{titlepage}

\renewcommand{\thesection}{\normalsize{\arabic{section}.}}
\vfill\eject

\section{Premise}

The discovery \cite{discovery1,discovery2} of the narrow scalar resonance with
mass $m_h=125$ GeV at the Large Hadron Collider (LHC) of CERN marked a
milestone in the field of particle physics. Extensive research has shown that
this boson couples to the other known particles proportionally to their
respective masses. Spontaneous symmetry breaking (SSB) through the Higgs field
was thus experimentally confirmed as the fundamental ingredient that fixes the
vacuum of electroweak interactions. 

But not everything may yet be fully understood. Indeed, within a perturbative
approach, the resulting scalar self-coupling $\lambda^{\rm (p)}(\phi)$
(p=perturbative) starts to decrease from its value
$\lambda^{\rm (p)}(v)=3m^2_h/v^2$ at the Fermi scale $v\sim 246$~GeV and
eventually becomes negative beyond an instability scale
$\phi_{\rm inst} \sim 10^{10}$~GeV. As a consequence, the true minimum of the
perturbative Standard Model (SM) potential would lie beyond the Planck scale
\cite{branchina,gabrielli} and be much deeper than the electroweak minimum.
This result, implying that the SM vacuum is a metastable state, requires
a cosmological perspective that raises several questions concerning the role
of gravity and/or the necessity to formulate the stability problem in the
extreme conditions of the early universe. The survival of the tiny
electroweak minimum is then somewhat surprising, which suggests that either
we live in a very special and exponentially unlikely corner or new physics
must exist below $\phi_{\rm inst}\sim 10^{10}$~GeV~\cite{riotto}.  

An alternative could be to first consider the pure scalar sector but describe
SSB consistently with studies indicating that the quadratic shape of the
potential at the minimum differs from the mass scale associated with the 
zero-point energy. Thus, the Higgs field could exhibit a second resonance
with a much larger mass, which stabilises the potential yet couples to
longitudinal $W$s just like the 125 GeV state and so remains a relatively
narrow resonance. In the present paper, we will first briefly summarise an
approach \cite{Cosmai2020}$ -$\cite{EPJC} that follows this line of thought
and predicts a second resonance of the Higgs field with the much larger mass
$(M_H)^{\rm Theor} = 690\,(30)$~GeV. For many details we will refer to
preceding articles, especially to the very complete analysis in
Ref.~\cite{EPJC}. Here, we have substantially improved upon our
phenomenological analysis.  Indeed, we include more LHC data, sharpen the
analysis of some final states, and indicate interesting correlations between
different channels that can only be explained with the existence of a new
resonance. Therefore, the combined statistical evidence, despite being roughly
estimated, could now be even above the traditional five-sigma level.

\section{A second resonance of the Higgs field}

By concentrating on a pure $\Phi^4$ theory, in
Refs.~\cite{Cosmai2020}$-$\cite{EPJC} a picture of SSB as a (weak)
first-order phase transition was adopted. This means that, as in the original
Coleman-Weinberg paper \cite{Coleman:1973jx}, SSB may originate from the
zero-point energy (ZPE) in the classically scale invariant limit
$ V''_{\rm eff}(\phi=0)\to 0^+$. A crucial point is that this description
is obtained in those Gaussian-like approximations to the effective potential
(one-loop potential, Gaussian effective potential, post-Gaussian calculations)
that encompass some classical background plus the ZPE of free-field-like
fluctuations with a $\phi$-dependent mass. In this sense, there is consistency
with the basic ``triviality'' of the theory in four dimensions (4D). This
first-order picture finds also support in lattice simulations. To that
end, one can just look at Fig.~7 in Ref.~\cite{akiyama2019phase},
where the data for the average field at the critical temperature
show the characteristic first-order jump and not a smooth
second-order trend.
\begin{figure}
\begin{center}
\includegraphics[bb=20 0 1100 500, angle=0,scale=0.2]{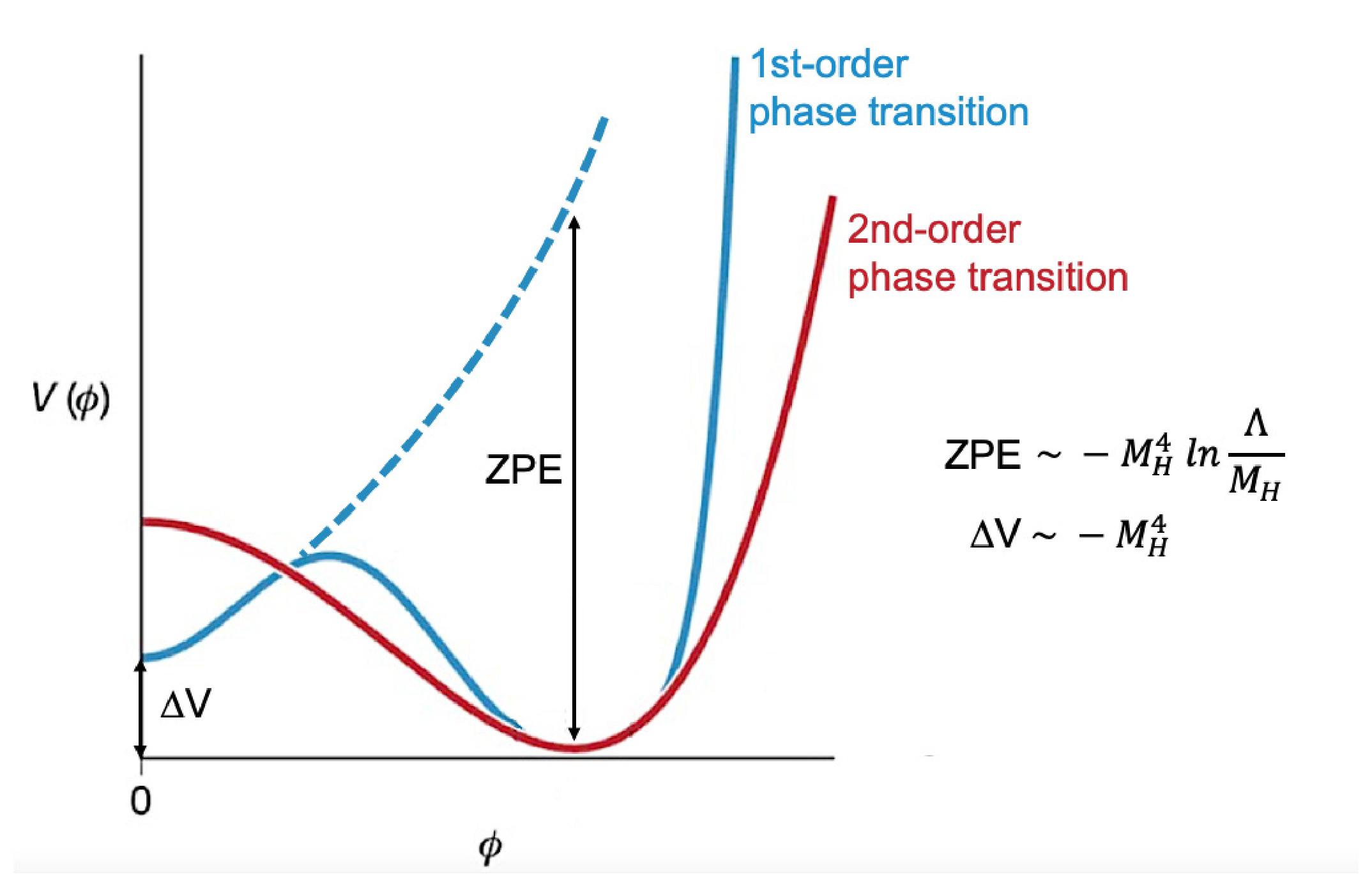}
\caption{\it An intuitive picture which illustrates the crucial role of the ZPE
in a first-order scenario of SSB. Differently from the standard second-order
picture, these have to compensate for a tree-level potential with no
non-trivial minimum.}
\end{center}
\label{first}
\end{figure}

At first sight, the nature of the phase transition may seem irrelevant,
because nothing prevents the potential from having locally the same shape
as in a second-order picture. To get more insight, let us look at Fig.~1.
This intuitively illustrates that, if $V''_{\rm eff}(\phi=0)> 0$, the ZPE
is expected to be much larger than in a second-order picture. In
the latter case, SSB is in fact driven by the negative mass-squared
at $\phi=0$, whereas now the ZPE has to overwhelm a tree-level potential
that otherwise would have no non-trivial minimum. Therefore, the ZPE mass
scale $M_H$ and the mass scale $m_h$, defined by the quadratic shape of
the effective potential at the minimum, could now be very different. 
Actually, a Renormalisation Group (RG) analysis of the effective potential
indicates that these two masses scale differently  with the ultraviolet
cutoff $\Lambda$.
Such an RG analysis is needed because, by ``triviality'', at any finite scale
$\mu$, the scalar self-coupling vanishes as
$\lambda(\mu)\!\sim \ln^{-1} (\Lambda/\mu)$, where $\Lambda$ is the Landau
pole fixing the cutoff scale. To minimise the cutoff dependence, one can thus
consider the whole set of theories ($\Lambda$,$\lambda$),
($\Lambda'$,$\lambda'$), ($\Lambda''$,$\lambda''$), \ldots, with larger and
larger cutoff values, smaller and smaller low-energy couplings at $\mu$, but
all sharing the same $\Lambda$-independent effective potential.
There are then two RG-invariant quantities, namely the mass scale $M_H$ itself
entering the minimum of the effective potential ${\cal E} \sim -M^4_H$ and a
particular definition of the vacuum field to be used for the Fermi scale
$v \sim 246$~GeV, which is always assumed to be cutoff independent. As such,
they can be related by some finite proportionality constant, say $M_H = K v$.
Instead, for the smaller mass $m_h$ defining the quadratic shape of the
potential, i.e., the inverse zero-momentum propagator $G^{-1}(p=0)$, one finds
$m^2_h \sim  M^2_H  L^{-1}\sim v^2 L^{-1} $  in terms of
$L=\ln (\Lambda/M_H)$, thus implying the traditional $\Phi^4$ relation
$\lambda(v)=3m^2_h/v^2\sim L^{-1}$. 

This mass structure was confirmed by explicit calculations of
the propagator from the corresponding Gaussian Effective Action (GEA)
\cite{okopinska}, both for the one-component and $O(N)$-symmetric theory,
with propagator 
\BE
\label{GEA_general}
G^{-1}(p)= p^2 +  M^2_H A(p) \; .
\EE
Indeed, upon minimisation of the Gaussian potential, this gives \cite{EPJC}
$G^{-1}_h(p)\sim  p^2 + m^2_h$ for $p \to 0$, where $A(p)\sim  L^{-1}$, and
$G^{-1}_H(p)\sim p^2 + M^2_H$ at larger $p^2$, where $A(p) \sim 1$.
The propagator structure in Eq.~(\ref{GEA_general}) was checked with lattice
simulations which are considered a reliable non-perturbative approach. These
simulations were also needed because the Gaussian-like approximations to the
effective potential that we have considered predict the same qualitative
scaling pattern but, resumming to all orders different classes of diagrams,
yield different values of the numerical coefficient $c_2$ controlling the
logarithmic slope, say $M^2_H \sim m^2_h L (c_2)^{-1}$. Therefore, using
numerical simulations \cite{Cosmai2020}, the best approximations to a
free-field propagator could be found and so compute $m_h$ from the
$p\to 0$ limit of $G(p)$, as well as $M_H$ from its behaviour at higher
$p^2$. In this way, the expected logarithmic trend was checked and $c_2$
extracted. Referring to Ref.~\cite{Cosmai2020,EPJC}, here we just summarise
the final result. The value $(c_2)^{-1/2} = 0.67\,(3)$ from the lattice was
replaced in the relation $M^2_H \sim m^2_h L (c_2)^{-1}$, so that by
combining with $m^2_h = \lambda (v) v^2/3$ and the leading-order trend
$\lambda(v)\sim (16 \pi^2/3)  L^{-1}$ of $\Phi^4$, the finite proportionality
relation $M_H = K v $ was obtained, with $K= (4\pi/3) (c_2)^{-1/2}$
or\footnote{Strictly speaking, $c_2$ was extracted from lattice
simulations of a one-component $\Phi^4$ theory. Thus, one could wonder about
the physical Higgs field described by an $O(4)$ theory. However, the effective
potential is rotationally invariant, so that basic properties of its shape,
such as the relation between the second derivative at the minimum and its
depth, should be the same as in a one-component theory. For a quantitative
argument, we recall that here one finds $m_h \ll M_H$ for very large
$\Lambda$. But $M_H$ is independent of $\Lambda$, so that by decreasing
$\Lambda$ the lower mass would increase and approach its maximun value
$(m_h)^{\rm max}\sim M_H \sim $ 690(30) GeV when $\Lambda$ becomes as small
as possible, say a few times $M_H$. If we then compare this prediction from
the one-component theory with the existing upper bounds from lattice
simulations of the $O(4)$ theory, we find a good consistency with
Lang's \cite{lang} and Heller's \cite{heller} values, viz.\
$(m_h)^{\rm max}=670\,(80)$~GeV and $(m_h)^{\rm max}=710\,(60)$~GeV,
respectively. Actually, the combination of these two estimates
$(m_h)^{\rm max} \sim 690\,(50)$~GeV would practically coincide with our
expectation. In this sense, we could have predicted the value of $M_H$
from these two old theoretical upper bounds without performing our own
lattice simulations of the propagator. At the same time, we should not
forget that in the real world $m_h=125$~GeV. Therefore, if there is a
second resonance with $M_H\sim 690$~GeV, the ultraviolet cutoff $\Lambda$
should be extremely large.}
\BE
\label{MHTHEOR}
(M_H)^{\rm Theor} = Kv= 690\,(30) \; {\rm GeV} \; .
\EE

\section{Basic phenomenological aspects }
\label{pheno}

The possible existence of a second, much larger mass $M_H\sim 690$~GeV
associated with the ZPE implies that the known gauge and fermion fields
would play a minor role for vacuum stability. In fact, by subtracting
quadratic divergences or using dimensional regularisation, the
logarithmically divergent terms in the ZPE due to the various fields are
proportional to the fourth power of the mass, so in units of the
pure scalar term one finds
$ (6 M^4_w + 3 M^4_Z)/M^4_H \lesssim 0.002$ and $12 m^4_t/M^4_H\lesssim 0.05$.
Besides, the two couplings $\lambda^{\rm (p)}(\mu)$ and $\lambda(\mu)$
coincide for $\mu=v$, so that their different evolution at large $\mu$
remains unobservable.
Confirming this alternative mechanism of SSB then requires the
observation of the second resonance and checking its phenomenology.

In this respect, the hypothetical $H$ is not like a standard Higgs boson of
700~GeV, as it would couple to longitudinal $W$s with the same typical
strength as the low-mass state at 125~GeV \cite{memorial,EPJC}. This can be
explicitly shown by the Equivalence Theorem, when understood as a
non-perturbative statement, valid to lowest non-trivial order in
$g^2_{\rm gauge}$ but also to all orders in the scalar self-couplings
\cite{bagger}. This way, in longitudinal $WW$ scattering the contact
coupling $\lambda_0= 3 M^2_H/v^2$, generated by the incomplete cancellation
of graphs at tree level, is transformed into
$\lambda(v)= 3 m^2_h/v^2= (m^2_h/M^2_H) \lambda_0$ after resumming graphs
to all orders. The equivalent argument is that it is $m_h= 125$~GeV, and not
$M_H\sim 700$~GeV which fixes the quadratic shape of the potential and the
interaction with the Goldstone bosons. 

Thus, the large conventional widths $\Gamma(H \to ZZ+WW)\sim G_F M^3_H$
would be suppressed by the small ratio $(m_h/M_H)^2\sim 0.032$,
leading to the estimates
$\Gamma(H \to ZZ)\sim \frac { M_H} { 700~ {\rm GeV}}\,\times\,$(1.60~GeV) and
$\Gamma(H\to WW)\sim \frac { M_H} { 700~ {\rm GeV}}\,\times\,$(3.27 GeV),
besides the new contribution
$\Gamma(H \to hh)\sim \frac { M_H} { 700~ {\rm GeV} }\,\times\,$(1.52 GeV). As
such, the heavy $H$ should be a relatively narrow resonance of total width
$\Gamma(H\to {\rm all}) =25\div35$~GeV, decaying predominantly to $t \bar t$
quark pairs, with a branching ratio of about 70$\div$80 $\%$.  Note the very
close branching ratios $B(H \to hh) \sim 0.95~ B(H \to ZZ)$. Finally, due to
its small coupling to longitudinal $W$s, $H$ production through vector-boson
fusion (VBF) would be negligible as compared to gluon-gluon fusion
(ggF), which has a typical cross section
$\sigma^{\rm ggF} (pp\to H) \sim 1100\,(170)$~fb \cite{widths,yellow},
depending on QCD and $H$-mass uncertainties.

\section{In touch with the experiments: ATLAS 4-lepton and
\boldmath{$\gamma\gamma$} data}
To get in touch with experiments, let us start from the four-lepton channel.
In a first approximation, resonant four-lepton production at the $H$ peak
could be estimated through the chain
\BE
\label{exp3}
\sigma_R \; \equiv \; \sigma_R (pp\to H\to 4l) \; \sim \; \sigma (pp\to H)
\times B( H \to ZZ) \times 4 B^2(Z \to l^+l^-)\;,
\EE
with $4 B^2(Z \to l^+l^-)\sim 0.0045$. Thus, by substituting
$\Gamma(H \to ZZ)\sim \frac { M_H} { 700~ {\rm GeV} }\,\times\,$(1.6 GeV),
$\Gamma(H \to {\rm all})=25\div 35$~GeV, and
$\sigma (pp\to H)\sim  \sigma^{\rm ggF} (pp\to H)\sim 1100\,(170)$~fb, we
would predict $\sigma_R \sim 0.26\,(7)$~fb.
\begin{figure}[ht]
\centering
\includegraphics[width=0.65\textwidth,clip]{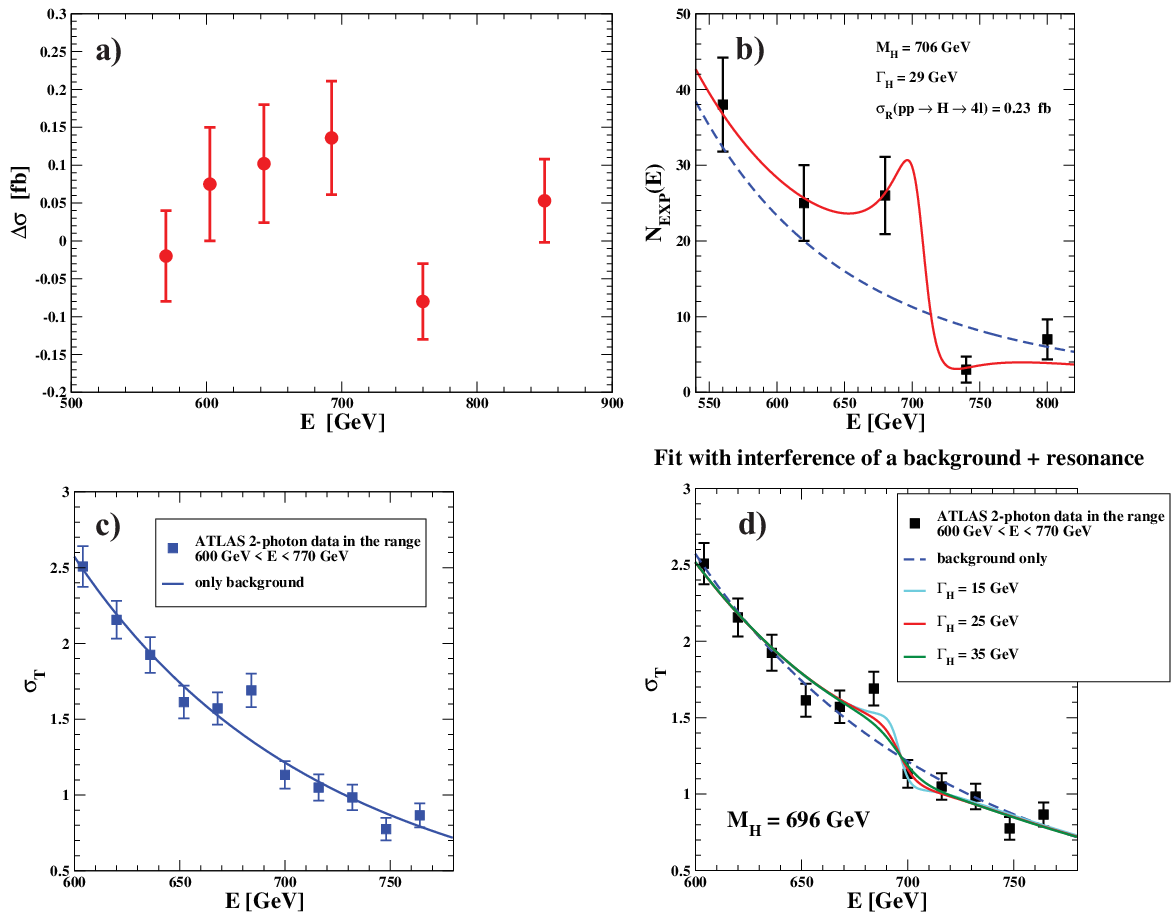}
\caption{\it Panel a) shows the cross-section difference $\Delta \sigma$
between the experimental data and the expected background, as measured by
ATLAS in the four-lepton channel \cite{atlasnew}. The numerical values and
energy bins can be found in Refs.~\cite{LHEP,EPJC}. Panel b) shows the number
of ATLAS ggF-low four-lepton events \cite{atlas4lHEPData}, grouped
into bins of 60~GeV from 530 to 830~GeV. The blue dashed curve is the
background while the red solid curve is the fit with Eq.~(\ref{sigmat}).
The numerical values and energy bins can be found in Refs.~\cite{LHEP,EPJC}.
Panel c) shows the invariant-mass
distribution of the inclusive $\gamma\gamma$ events (converted to
cross sections in fb), observed by ATLAS \cite{atlas2gammaplb} in the range 
$\mu(\gamma\gamma)=E= 600\div 770$~GeV and fitted with background only.
The numerical values and energy bins can be found in Refs.~\cite{LHEP,EPJC}.
Finally, panel d) shows the fit to the data with Eq.~(\ref{sigmat}),
$M_H=696$~GeV, and three values of $\Gamma_H$.}
\label{4figures}
\end{figure}
In the previous analysis of Refs.~\cite{LHEP,EPJC}, a comparison was made to
the data in Fig.~\ref{4figures}. Its panel a) reports the cross-section
difference $\Delta \sigma$ between the experimental data and the expected
background, as measured by ATLAS in the four-lepton channel \cite{atlasnew}.
There is an excess-defect pattern that may indicate the characteristic change
of sign of the interference past a Breit-Wigner peak. The numerical
$\Delta \sigma$ and the size of the bins are given in Refs.~\cite{LHEP,EPJC}.
Here, for convenience of the reader, we just report the $\Delta \sigma$ for
the four central bins from 585 to 800~GeV, whose individual values in fb are
$0.085\pm0.075$, $0.102\pm0.078$, $0.136\pm0.075$, and $-0.080 \pm 0.050$,
respectively. To describe these data, we have adopted the model of a resonance
that interferes with a given background $\sigma_b(E)$, giving rise to a total
cross section ($s=E^2$ and $\Gamma_H=\Gamma(H\to {\rm all})$)
\BE
\label{sigmat}
\sigma_T(E) \; = \; \sigma_b(E) + \sigma_R \cdot R(E) + \sigma_{\rm int}(E)\;,
\EE
where
\BE
\sigma_{\rm int}(E) \; = \; 2\sqrt{\sigma_b(E)\sigma_R} \,
\frac{(M^2_H-s) }{\Gamma_H M_H} \, R(E)
\EE
and
\BE
\label{Rs}
R(E) \;  = \; \frac{(\Gamma_H M_H)^2 }{(s-M^2_H)^2+(\Gamma_H M_H)^2} \; .
\EE
Some refinement is needed if one assumes the resonance to be produced through
a specific parton process, e.g.\ through ggF as in our case. To implement this
refinement and denoting as $\sigma^{\rm gg}_b$ the specific four-lepton
background cross section from the ggF mechanism given in Ref.~\cite{atlasnew},
the ``non-ggF'' background was preliminarily subtracted in
Ref.~\cite{LHEP,EPJC} by defining a modified experimental cross section
\BE
\label{sigmahat}
\hat{\sigma}_{\rm EXP} \; = \; \sigma_{\rm EXP} - (\sigma_b-\sigma^{\rm gg}_b)
\EE
and then replacing everywhere $\sigma_b \to \sigma^{\rm gg}_b$ in the theoretical
Eq.~(\ref{sigmat}).  The thus fitted values were $M_H= 677^{+30}_{-14}$~GeV,
$\Gamma_H= 21^{+28}_{-16}$~GeV, and $\sigma_R= 0.40^{+0.62}_{-0.34}$~fb.
As an additional check, we also considered \cite{universe,LHEP,EPJC} the other data in panel b) of
Fig.~\ref{4figures}. This shows the statistically dominant ggF-low sample
of ATLAS four-lepton events \cite{atlas4lHEPData}, grouped into bins of
60~GeV from 530 to 830~GeV. Fitting this other set of data gave similar
results,  viz.\ $M_H=706\,(25)$~GeV,  $\Gamma_H= 29\pm 20$~GeV, and
$\sigma_R= 0.23^{+0.28}_{-0.17}$~fb. However,  this tentative agreement
reflects the rather large error bars of the data. In fact, these events
include a sizeable contribution from $q \bar q \to ZZ \to 4l $ processes
that, strictly speaking, should not interfere with a resonance solely
produced through the ggF mechanism. 
In any case, our expected mass  $(M_H)^{\rm Theor} = 690\,(30)$~GeV and width
$\Gamma_H= 25\div 35$~GeV were well consistent with both types of fit.

Looking for other indications, the invariant-mass distribution of the
inclusive $\gamma\gamma$ events observed by ATLAS \cite{atlas2gammaplb} in
the range $\mu(\gamma\gamma)=E= 600\div 770$~GeV was also considered
\cite{universe,LHEP,EPJC}. By parametrising the background with a power-law
form $\sigma_b(E) \sim A\cdot ({\rm 685~GeV}/E) ^{\nu}$ one gets a good
description of all data points, except for the sizeable excess at 684~GeV,
which was estimated by ATLAS to have a local significance of more than
$3\sigma$ (see panel c) of Fig.~\ref{4figures}). This isolated discrepancy
shows how a new resonance might remain hidden behind the large background
nearly everywhere. For this reason, by fitting to Eq.~(\ref{sigmat}), with
the exception of the mass $M_H=$ 696\,(12) GeV, the total decay width was
determined very poorly, namely $\Gamma_H= 15^{+18}_{-12}$~GeV. In panel d)
of Fig.~\ref{4figures} we report three fits for $M_H= 696$~GeV and
$\Gamma_H=$ 15, 25, and 35~GeV, respectively.

Before concluding this section, two considerations are in order. First, with
a definite prediction $(M_H)^{\rm Theor}=690\,(30)$~GeV, one should look for
deviations from the background nearby, say in the mass region
600$\div 800$~GeV, so that local deviations should {\it not} \/be downgraded
by the so-called ``look elsewhere'' effect. Secondly, the local statistical
significance of deviations from the background should take into account the
phenomenology of a resonance that can produce {\it both} \/excesses and
defects of events. For this reason, the statistical significance of the
deviations from background seen in panel a) of Fig.~\ref{4figures} is
actually $3\sigma$, like for the $\gamma\gamma$ data in panel c).  

\section{The CMS 4-lepton events}

\begin{table*}
\caption{\it We specify in the first three columns: the four-lepton invariant
mass $m(4l)=E$, the expected CMS background events, and the experimental $S/B$
ratio reported in Fig.~7, upper left panel left, of
Ref.~\cite{CMS_4leptons_2024}, respectively. We then present the theoretical
value of Eq.~(\ref{SBratio}) for the optimal set of parameters obtained in
the fit,  viz.\ $M_H=692$~GeV, $\Gamma_H=10$~GeV, $N_R=0.55$, and the
chi-squared of the fit. The numerical values of background events and
experimental $S/B$ ratio, not reported in Ref.~\cite{CMS_4leptons_2024},
were directly extracted from the figures. The accuracy is about 3$\div 4$\%.}
\begin{center}
\begin{tabular}{ccccc}
$\rm E$[GeV] & $N _b(E)$ &~[S/B]$^{\rm EXP}$
&~~~ [S/B]$^{\rm Theory}$ & $\chi^2$\\
\hline \hline
645& 1.46\,(6) & 1.10 (42) & 1.14 & 0.01  \\
\hline
660& 1.33\,(5) & 1.20 (45) & 1.21 &0.00  \\
\hline
675 & 1.20\,(5) & 1.56 (58) & 1.41& 0.07 \\
\hline
690 & 1.09\,(4) & 1.93 (67)  & 1.88 & 0.01 \\
\hline
705 & 0.99\,(4) & 0.54 (38)  & 0.58 &0.01 \\
\hline
720 & 0.90\,(4) & 1.19 (61) & 0.76 & 0.50 \\
\hline
735 &0.82\,(3) & 0.98 (57) & 0.83 & 0.07 \\
\hline
\end{tabular}
\end{center}
\label{CMSleptontable}
\end{table*}

\begin{figure}[ht]
\centering
\includegraphics[width=0.40\textwidth,clip]{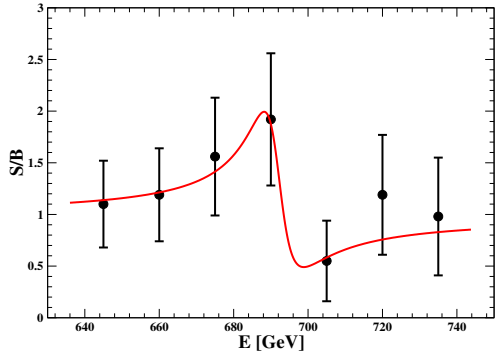}
\caption{\it The fit with Eq.~(\ref{SBratio}) to the data in
Table~\ref{CMSleptontable} for $M_H=692$~GeV, $\Gamma_H=10$~GeV, and
$N_R=0.55$. Note the close similarity with panels a) and b) of
Fig.~\ref{4figures}.}
\label{CMSvecchiofit}
\end{figure}
We will now compare with the recent, still preliminary CMS data for the
four-lepton channel in Ref.~\cite{CMS_4leptons_2024}. To this end, we will
first transform from Eq.~(\ref{sigmat}) to the number of events $N$, for
a given luminosity and acceptance, thus finding a total number 
\BE
N_{\rm T} \; = \; N_b(E) +  2 \sqrt{ N_b(E) N_R } \,
\frac{(M^2_H-s) }{\Gamma_H M_H} \, R(E) + N_R R(E)
\EE
and a theoretical $S/B$ ratio 
\BE
\label{SBratio}
[S/B]^{\rm Theory} \; = \; 1 + 2 \sqrt{\frac{N_R}{ N_b(E)}} \,
\frac{(M^2_H-s)}{\Gamma_H M_H} \, R(E) + \frac{N_R}{ N_b(E)} \, R(E) \; .
\EE
The CMS data for expected background events $N_b(E)$ and experimental
$S/B$ ratio are given in Table~\ref{CMSleptontable}. The combined deviation
from unity of the three points at 675, 690, and 705 GeV is $2\sigma$. A fit
to these data with Eq.~(\ref{SBratio}) yields $M_H= 692^{+17}_{-12}$~GeV,
$\Gamma_H= 10^{+26}_{-8}$~GeV, and $N_R= 0.55^{+5.0}_{-0.45}$. The predictions
of Eq.~(\ref{SBratio}) for the optimal parameters are also presented in
Table~\ref{CMSleptontable} and a graphical comparison is shown in
Fig.~\ref{CMSvecchiofit}. 
\begin{figure}[ht]
\centering
\includegraphics[width=0.50\textwidth,clip]{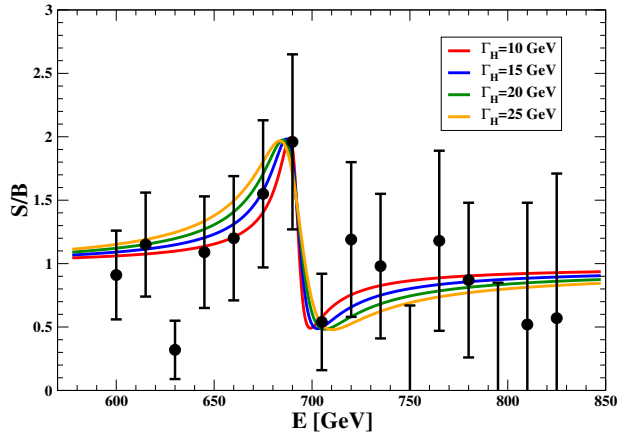}
\caption{\it Comparing the CMS $S/B$ ratios to Eq.~(\ref{SBratio}),
for $M_H=692$~GeV, $N_R=0.55$, and four different widths.}
\label{CMSnuovofit}
\end{figure}

To have a more complete idea of the overall agreement with the CMS data, we
also enlarge the energy range from 600 to 800~GeV. The data for the
$S/B$ ratio are then presented in Fig.~\ref{CMSnuovofit}, together with
various curves for the same pair $M_H= 692$~GeV, $N_R=0.55$ and different
values of $\Gamma_H$. The more refined treatment of subtracting preliminarily
the non-ggF background and comparing with the modified Eq.~(\ref{sigmahat})
is not possible here, in view of the very large error bars of the data.  

\section{CMS-TOTEM \boldmath{$\gamma\gamma$} events produced in
\boldmath{$pp$} diffractive scattering}

\begin{figure}[ht]
\centering
\includegraphics[width=0.65\textwidth,clip]{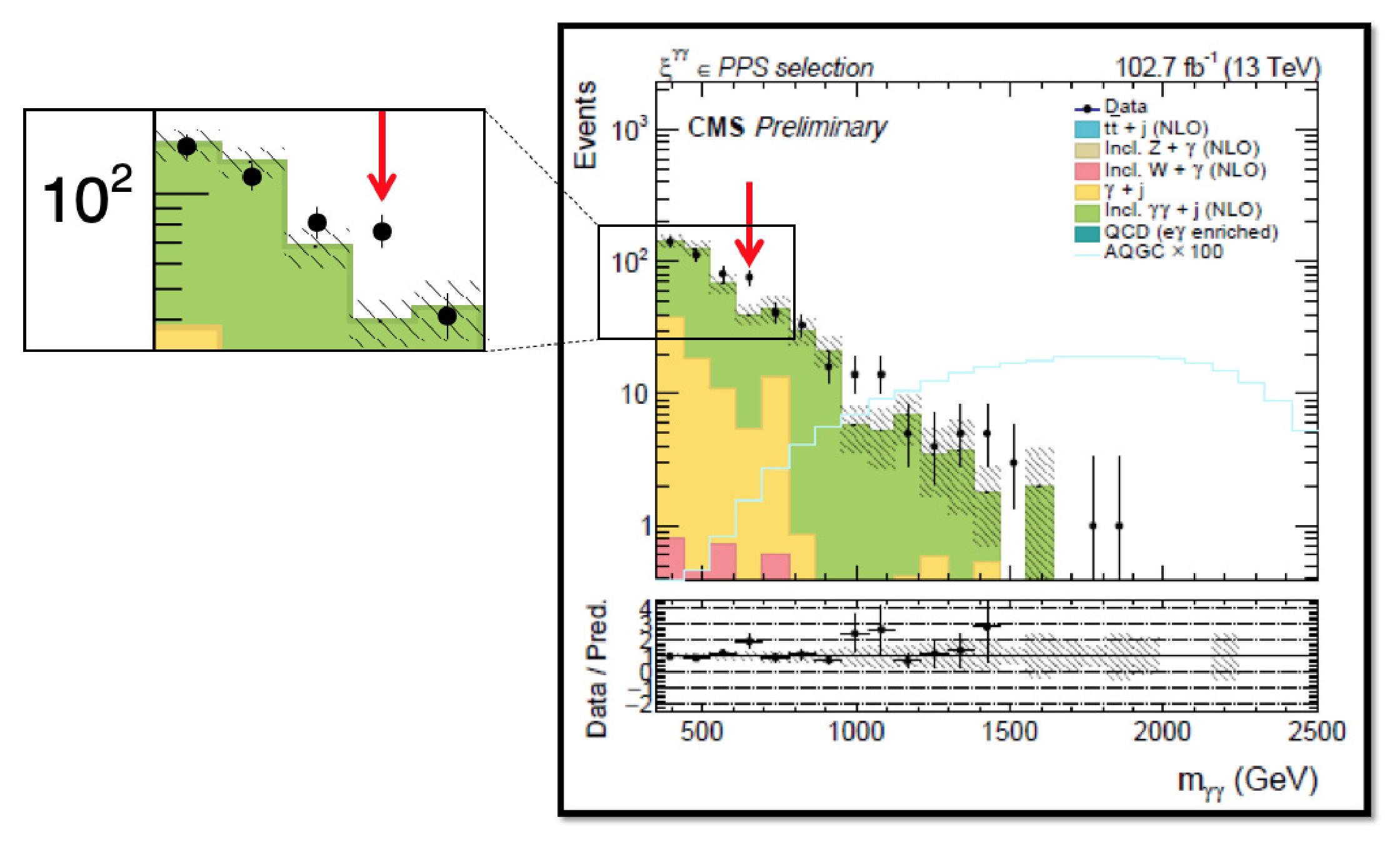}
\caption{\it The number of $\gamma\gamma$ pairs produced in $pp$
diffractive scattering as reported in
Ref.~\cite{PRD_CMS_TOTEM}. In the range $650\,(40)$~GeV, the observed
number is $N_{\rm exp}\sim 76\,(9)$, to be compared to an estimated
background $N_{\rm bkg}\sim 40\,(9)$.}
\label{diffractive}
\end{figure}

The CMS and TOTEM Collaborations have also been searching for
high-mass photon pairs produced in $pp$ double-diffractive
scattering, i.e., when both final protons are tagged and have large
$x_F$. For our purpose, the relevant information is contained in
Fig.~\ref{diffractive} taken from Ref.~\cite{PRD_CMS_TOTEM}. In the
range of invariant mass $650\,(40)$~GeV, and for a statistics of
102.7~fb$^{-1}$ the observed number of $\gamma\gamma$ events was
$N_{\rm exp}\sim 76\,(9)$, to be compared with an estimated background
$N_{\rm bkg}\sim 40\,(9)$. In the most conservative case,
viz.\ $N_{\rm bkg}=49$, this represents a local $3\sigma$ effect and is
the only statistically significant excess in the plot.

\section{The ATLAS \boldmath{$t \bar t$} events}

The ATLAS Collaboration also searched for scalar resonances decaying to
top-quark pairs \cite{ATLAS_conf}. There are small excesses at
$675\,(75)$~GeV in the invariant mass of the $llbb$ system, which are more
evident when the tracks of the final leptons are at large angles. The excess
is minuscule, because the expected signal for a 700~GeV Higgs is about 1~pb,
to be compared with a background cross section of $107.0\,(7.6)$~pb
(see the CMS measurement \cite{CMS_top} of top-quark pairs for invariant
mass 620 $\div$ 820~GeV). 
\begin{figure}[ht]
\centering
\includegraphics[width=0.35\textwidth,clip]{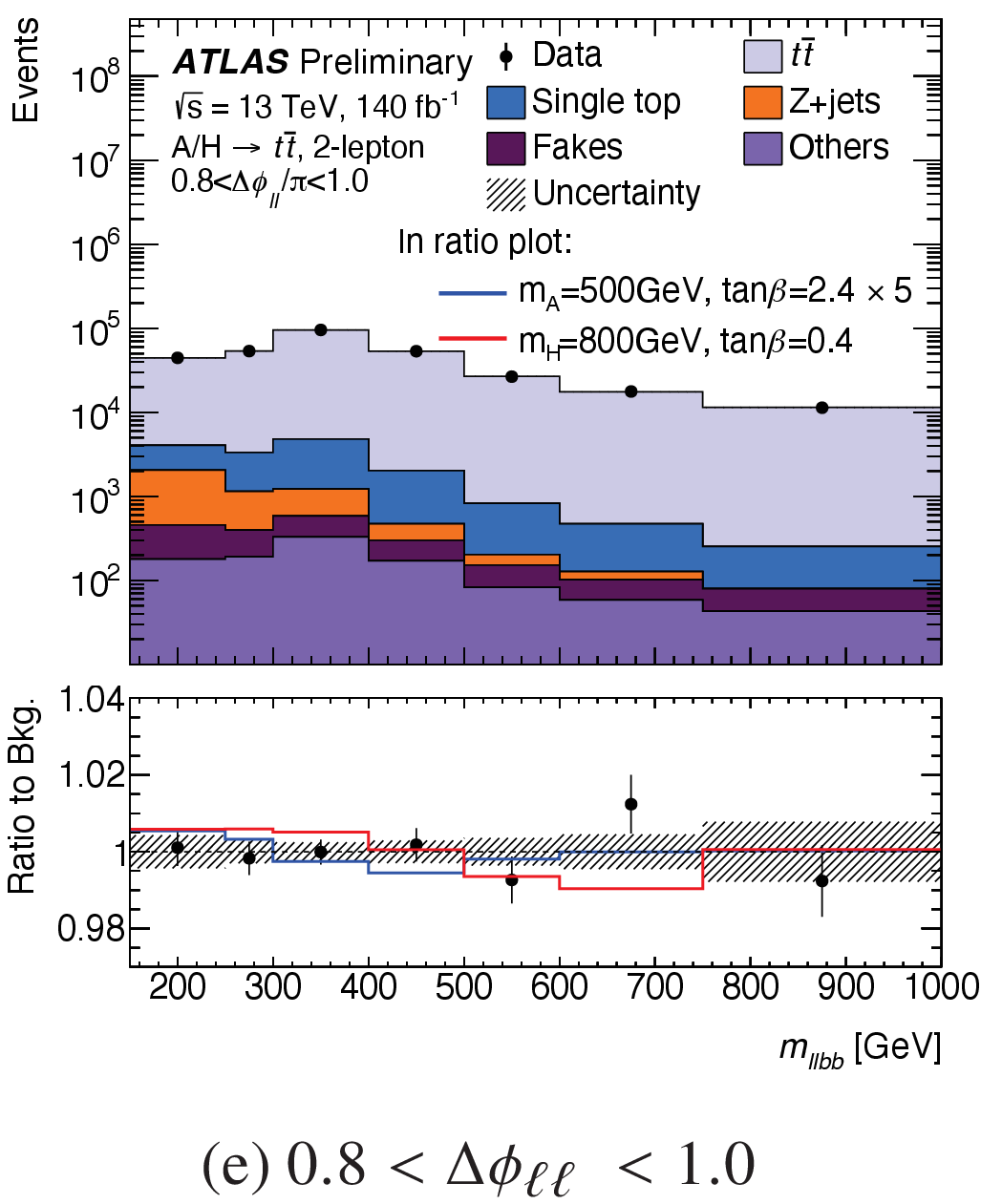}
\caption{\it The slight excess of $t \bar t$ pairs observed by ATLAS
for an invariant mass of the $llbb$ system around 675~GeV. }
\label{ttbar}
\end{figure}

\section{A closer look at the  ATLAS \boldmath{$b\bar b+\gamma\gamma$} data }

\begin{table*}
\caption{\it For the bins $550\,(25) \div 800\,(25)$~GeV in
Fig.~\ref{ATLAS_BBGG}, we report the experimental 95\% upper limits
$\sigma^{\rm exp}(j)$ for the cross section $\sigma(pp\to X \to h h)$
(black dots). The index $j=1\ldots6$ indicates the bins
$550\,(25), 600\,(25), \ldots, 800\,(25)$~GeV, respectively. Error bars in
the experimental entries only take into account the $\sqrt{N}$ statistical
uncertainty of the final $b\bar b + \gamma\gamma$ events. In the third
column, we give the expected background values with $\pm1\sigma$ and
$\pm2\sigma$ uncertainties (see the HEPData file of
Ref.~\cite{ATLAS_BBGG_paper}).}
\begin{center}
\begin{tabular}{ccc}
\hline\hline
j & $\sigma^{\rm exp}$(j) [fb] & $\sigma^{\rm bkg}$(j) [fb] \\
\hline
1 & 87.5\,(15.6) &   $95.1^{ {+50.4}^{+137.3} }_ { {-26.6}_{-44.1} } $  \\
\hline
2 & 73.6\,(14.3) & $81.1^{ {+43.3}^{+119.0} }_ { {-22.7}_{-37.6} } $    \\
\hline
3 & 149.3\,(20.3) & $84.4^{ {+44.4}^{+120.1} }_ { {-23.6}_{-39.1} } $  \\
\hline
4 & 49.4\,(12.0)  & $76.5^{ {+40.0}^{+109.6} }_ { {-21.4}_{-35.4} } $  \\
\hline
5 & 44.5\,(12.0)  & $71.7^{ {+37.6}^{+103.3} }_ { {-20.0}_{-33.2} } $\\
\hline
6 & 71.0\,(14.0)   &  $65.8^{ {+35.1}^{+96.5} }_ { {-18.4}_{-30.5} } $\\
\hline
\end{tabular}
\end{center}
\label{sigmas}
\end{table*}

\begin{figure}[htb]
\includegraphics[width=0.55\textwidth,clip]{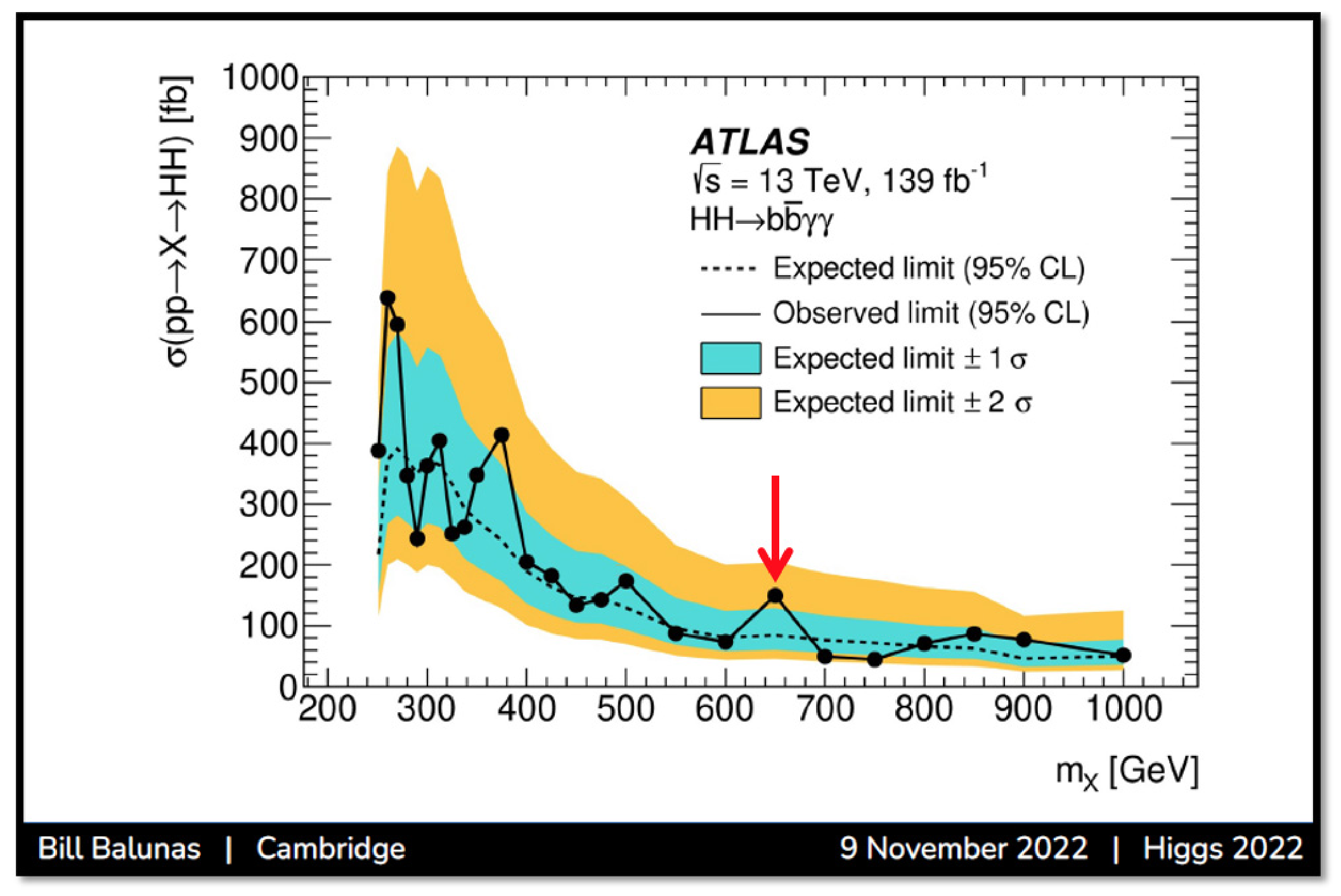}
\caption{\it Expected and observed 95$\%$ upper limit for the cross
section $\sigma (pp\to X \to h(125)h(125))$ obtained by ATLAS
\cite{ATLAS_BBGG_paper} from the final state $(b \bar b + \gamma\gamma)$.
The figure is taken from the talk given by Bill Balunas at ``Higgs 2022''
and is the same as Fig.~15 in Ref.~\cite{ATLAS_BBGG_paper}.}
\label{ATLAS_BBGG}
\end{figure}

The ATLAS Collaboration has searched for a new resonance $X$ decaying,
through a pair of $h=h(125)$ scalars, into the  $b\bar b+\gamma\gamma$
final state \cite{ATLAS_BBGG_paper}. Their results in Fig.~\ref{ATLAS_BBGG}
are given in terms of 95$\%$ upper limits for the cross section
$\sigma (pp\to X \to hh)$, as a function of the invariant mass of the
$b\bar b+\gamma\gamma$ system. The measured values, say 
$\sigma^{\rm exp}(j)$ in each bin $j$ (the black dots), are then compared
with the expected limits, say  $\sigma^{\rm bkg}$(j), along the black dashed
line, by also allowing for $\pm1\sigma$ and $\pm2\sigma$ uncertainties in
the theoretical predictions (see Table~\ref{sigmas}). 

To compare with a resonance around 690~GeV, we will restrict ourselves to
the mass range $550\div800$~GeV. At first sight, this indicates a modest
excess at $650\,(25)$~GeV, followed by two slight defects at 700 and 750~GeV.
Such an excess-defect pattern could indicate the interference with the
background around a Breit-Wigner peak and, in this interpretation, the
mass would lie between 650 and 700~GeV, say $M_H \sim 675\,(25)$~GeV,
where the interference changes sign.
Now, the process $pp\to hh \to b \bar b + \gamma\gamma$ has
also background contributions that should give no interference with a
resonance solely produced by the ggF mechanism. Since differently
from the four-lepton channel, the pure ggF contribution to the background
$\sigma^{\rm gg}_b$ is not given explicitly here, one cannot adopt the most
accurate procedure of first subtracting the non-ggF background and compare
with Eq.~(\ref{sigmahat}). Besides, the ATLAS entries in Table~\ref{sigmas}
express upper bounds, so that, strictly speaking, one cannot fit with
Eq.~(\ref{sigmat}) to extract $M_H$ and $\Gamma_H$. Nonetheless, one can try
to understand the order of magnitude of the main effect: the very
large difference between the two entries at 650 and 700 GeV.
To this end, let us assume a mass value $M_H \sim 675$~GeV. With the numerical
values in Sec.~3, we then expect a peak cross section
$\sigma_R=\sigma(pp\to H) B(H\to hh) \sim 55\,(10)$~fb,
whose uncertainty comes mainly from the ggF production cross section, because both
the partial and total decay widths scale linearly with mass. By also assuming
$\Gamma_H \sim 25$ GeV and the same central values for the background as in the
third column of Table~\ref{sigmas}, from Eq.~(\ref{sigmat}) we would then expect
the pair $\sigma_T(650) \sim 150$~fb and $\sigma_T(700) \sim 37$~fb, which lie
very close to the experimental values

However, this is only a first level of comparison with these data.
Our point is that the modest statistical consideration, given so far to
the ATLAS $ b\bar b+\gamma\gamma$ data, was substantially influenced by the
large uncertainty in the expected limits, as given by the wide blue and
yellow bands around the central dashed line in Fig.~\ref{ATLAS_BBGG}. The
uncertainty in each absolute value of the cross sections is indeed large, but
this is not the right perspective. In fact, in our mass region and to a very
good approximation, the whole effect of these uncertainties is simply to
shift the line of the central values up and down. For instance, at 650~GeV
the experimental value $149.3\,(20.3)$~fb is well within the $+2\sigma$ limit
for the background $84.4 + 120.1 = 204.5$~fb. But if we now evaluate the
difference between the experimental values at 650 and 600~GeV, which is
$149.3\,(20.3) - 73.6\,(14.3) = 75.7\,(24.8)$~fb, this is much larger than
the corresponding background differences, either along the black central
line $84.4-81.1=3.3$~fb or along the $1\sigma$ and $2\sigma$ contours, being
$128.8 - 124.4=4.4$~fb and $204.5 - 200.1= 4.4$~fb, respectively. That would
now give a discrepancy of about $2.9\sigma$. 

Therefore, we have done the exercise of comparing the experimental differences
in consecutive energy bins
\BE
\Delta^{\rm exp}(j+1,j) \; = \; \sigma^{\rm exp}(j+1) -\sigma^{\rm exp}(j)
\EE 
 with the corresponding expected values
$\Delta^{\rm bkg}(j+1,j)=\sigma^{\rm bkg}(j+1)-\sigma^{\rm bkg}(j)$, which
remain nearly constant when evaluated on the black line or along the
corresponding boundaries of the blue and yellow bands; see Table~\ref{diff}. 
\begin{table*}
\caption{\it The experimental difference
$\Delta^{\rm exp}(j+1,j)=\sigma^{\rm exp}(j+1)-\sigma^{\rm exp}(j)$ in fb, 
as computed from the values in Table~\ref{sigmas}. We also report the expected
backround values $\Delta^{\rm bkg}(j+1,j)$. The latter are computed  on the
black dashed line and along the corresponding boundaries of the blue and
yellow bands.}
\begin{center}
\begin{tabular}{cc}
\hline\hline
$\Delta^{\rm exp}(2,1)= -13.9\pm 21.1$ &
$\Delta^{\rm bkg}(2,1) = -19.9 \pm 12.4$\\
\hline
${ \Delta^{  {\rm {exp} }   }(3,2)}= { +75.7 \pm 24.8 }$ &
${\Delta^{  {\rm { bkg} }   }(3,2)}= { +3.1 \pm 1.4 }$    \\
\hline
${ \Delta^{  {\rm { exp} }   }(4,3)}= { -99.9 \pm 23.6 }$ & 
${\Delta^{  {\rm { bkg} }   }(4,3)}= { -11.4 \pm 7.2 }$    \\
\hline
$\Delta^{\rm exp}(5,4)= -4.9 \pm 17.0 $ &
$\Delta^{\rm bkg}(5,4)= -6.8\pm 4.2 $ \\
\hline
${ \Delta^{  {\rm {exp} }   }(6,5)}= { +26.5 \pm 18.4 }$ &
${\Delta^{  {\rm {bkg} }   }(6,5)}= {-7.9 \pm 4.7 }$    \\
\hline
\end{tabular}
\end{center}
\label{diff}
\end{table*}
Looking at Table~\ref{diff}, the results are seen to indicate that the large
difference between the fourth and third experimental entries, viz.\
$-99.9 \pm 23.6$~fb, cannot be explained by theoretical uncertainties, which
would rather predict a difference in the range $-11.4 \pm 7.2 $~fb. Here, the
discrepancy is about $3.4\sigma$ and goes in the opposite direction. By also
including the discrepancy of $1.6\sigma$ between the fifth pair of entries,
the combined deviations reach the level of about $3.8\sigma$.

Other interesting observables are the measured ratios 
\BE
R^{\rm exp}(j,j+1) \;=\; \frac{\sigma^{\rm exp}(j)}{\sigma^{\rm exp}(j+1)} \;,
\EE 
because systematic effects, as many of those reported in Table VIII of 
Ref.~\cite{ATLAS_BBGG_paper} and which modify the overall normalisation of
the data, would cancel out. We have thus compared in Table~\ref{ratios} with
the corresponding background quantities
$R^{\rm bkg}(j,j+1)= \sigma^{\rm bkg}(j)/\sigma^{\rm bkg}(j+1)$. 
From the $4\sigma$ difference in the second row and the $2.2\sigma$ difference
in the fifth row, this would now give a combined value of about $4.5\sigma$. 
However, there is some ambiguity here, because by replacing $j \to j+1$ and
$j+1 \to j$ in Table~\ref{ratios} error bars now become asymmetric and the
individual deviations are not the same. This ambiguity is not present in the
$\Delta$s, because with the replacements $j \to j+1$ and $j+1 \to j$ there is
only a change of sign and the statistical significance of any deviation
remains the same. For this reason, we will limit ourselves to consider the
deviations observed in the $\Delta$s.

\begin{table*}
\caption{\it The experimental ratios
$R^{\rm exp}(j,j+1)=\sigma^{\rm exp}(j)/\sigma^{\rm exp}(j+1)$, as computed
from the values in Table~\ref{sigmas}. We also report the expected background
values of $R^{\rm bkg}(j,j+1)$. The latter are computed at the points on the
black dashed line for each pair of bins $j$ and $j+1$, as well as along the
corresponding boundaries of the blue and yellow bands. }
\begin{center}
\begin{tabular}{cc}
\hline\hline
$R^{\rm exp}(1,2)= 1.19(31)$ & $R^{\rm bkg}(1,2) = 1.18(1)$\\
\hline
${R^{  {\rm { exp} }   }(2,3)}= { 0.49(12) }$ &
${ R^{  {\rm { bkg} }   }(2,3)}= { 0.97(1) }$    \\
\hline
${ R^{  {\rm { exp} }   }(3,4)}= {3.02(84) }$ &
${ R^{  {\rm { bkg} }   }(3,4)}= { 1.10(1) }$    \\
\hline
$R^{\rm exp}(4,5)= 1.11(38)$ &
$R^{\rm bkg}(4,5) = 1.07(1) $ \\
\hline
${ R^{  {\rm { exp} }   }(5,6)}= { 0.63(21) }$ & 
${ R^{  {\rm { bkg} }   }(5,6)}= { 1.09(1) }$    \\
\hline
\end{tabular}
\end{center}
\label{ratios}
\end{table*}

Before concluding, we observe an interesting correlation. From the two ATLAS
bins at 700 and 750~GeV one finds a ratio $R^{\rm exp}(4,5) \sim 1.11$ that
nearly coincides with the corresponding background value
$R^{\rm bkg}(4,5) \sim 1.07 $. This is because the two values in
Table~\ref{sigmas}, viz.\ $\sigma^{\rm exp}(4) \sim 49.4$~fb and
$\sigma^{\rm exp}(5) \sim 44.5$~fb, while considerably smaller than the
corresponding average background values
$\langle \sigma^{\rm bkg}(4) \rangle= 76.5$~fb and
$\langle \sigma^{\rm bkg}(5)\rangle = 71.7$~fb, give the same average
$\langle S/B \rangle\sim$ 0.63. A possible explanation can be obtained by
looking at the CMS data for the $S/B$ in Fig.~\ref{CMSnuovofit}. This shows
that the bins at 750 and 795~GeV are empty, with their error bars representing
the CMS estimates for the upper limits which one could expect with more
statistics, about 0.66 and 0.86, respectively. Note that the first upper bound
at 750~GeV is only slightly higher than the lower bound, about 0.58, obtained
from the bin at 720~GeV. In view of the large error bars of the remaining
points, this means that values with $S/B$ considerably smaller than unity have
a large probability content. The theoretical curves, especially those of green
and yellow colour for widths 20$\div$25 GeV, can thus provide a clue
with their prediction of a slow increase in $S/B$ from about 0.5 at 705~GeV
up to about 0.8 at 800~GeV, with an average value $S/B \sim  0.65\,(15)$.
Still focusing on the four-lepton channel, let us return to panel a) of
Fig.~\ref{4figures} and to the difference $\Delta \sigma=-0.080\,(50)$~fb
between the average cross section
$\langle \sigma^{\rm exp}(4l)\rangle =0.126\,(47)$ fb measured by ATLAS in
the range  $720\div 800$~GeV and the corresponding expected background
$\langle \sigma^{\rm bkg}(4l)\rangle=0.206\,(18)$~fb; see Table 4 of
Ref.~\cite{EPJC}. From the ratio of these two cross sections, we thus obtain
the average ratio $S/B = 0.61\,(23)$ measured by ATLAS and, in view of its
consistency with the previous value $0.65\,(15)$, an average combined
$\langle S/B \rangle^{\rm 4l} = 0.64\,(13)$ from the four-lepton channel past
the resonance peak. Since $M_H$ and $\Gamma_H$ are the same for both $
pp\to H\to ZZ$ and $pp\to H \to hh$, and the two branching ratios
$B(H\to hh)$ and $ B(H\to ZZ)$ are very close, we can use this combined
value to describe the analogous reduction of events observed in the
$b \bar b +\gamma\gamma$ final state.
The predicted averages
\BE
\langle \sigma(4)\rangle \sim\langle S/B \rangle^{\rm 4l} \langle
\sigma^{\rm bkg}(4)\rangle \;=\; 49.0\,(9.9)\:{\rm fb} \;\;\; {\rm and} \;\;\;
\langle \sigma(5)\rangle \sim \langle S/B \rangle^{\rm 4l} \langle
\sigma^{\rm bkg}(5) \rangle \;= \; 45.9\,(9.3)\:{\rm fb}
\EE
are then in very good agreement with the experimental values in
Table~\ref{sigmas}, confirming at the same time the accuracy of the average
background estimates. The existence of this correlation, between four-lepton
and  $b \bar b +\gamma\gamma$ final states, could hardly be explained without
the second resonance.

Summarising: in the region of invariant mass that is crucial for the
predicted second resonance $H$ of the Higgs field, the ATLAS determinations
of the cross section $\sigma^{\rm exp}(pp \to X \to h h)$ from the
$b \bar b +\gamma\gamma$ channel \cite{ATLAS_BBGG_paper} exhibit the same
characteristic excess-defect pattern observed by ATLAS and CMS in the
four-lepton channel. The natural interpretation is in terms of a resonance
with mass $M_H \sim 675\,(25)$~GeV. To estimate precisely the statistical
significance of the measurements, we have considered the differences of the
$\sigma^{\rm exp}(j)$ in consecutive bins and compared with the corresponding
combinations of the $\sigma^{\rm bkg}(j)$, where all theoretical uncertainties
nearly vanish. The combined statistical significance of the observed
deviations could then be estimated at the level of about $3.8\sigma$.
Of course, this is the statistical significance with our experimental
error bars (always larger than the size $\pm 12$~fb of the black dots in
Fig.~\ref{ATLAS_BBGG}), which only take into account the statistical
uncertainty in the determinations of the final $b\bar b+\gamma\gamma$ events.
On the other hand, while it is true that no other source of uncertainty is
included, systematic effects, as many of those reported in Table VIII of
Ref.~\cite{ATLAS_BBGG_paper} and affecting the overall normalisation of the
data, would cancel out in the ratios that also exhibit large deviations. 

\section{ Summary and conclusions}
In the present paper, we have first briefly summarised an alternative picture
of SSB, which predicts a relatively narrow second resonance of the Higgs
field, with mass $(M_H)^{\rm Theor} = 690\,(30)$~GeV. We then started
to compare with the LHC data.
Here, one should take into account three aspects that characterise this
particular research. First, with a definite prediction
$(M_H)^{\rm Theor}=690\,(30)$~GeV, one should look for deviations from the
background nearby, say in the mass region 600$\div 800$~GeV, so that local
deviations {\it cannot} \/be downgraded by the so called ``look elsewhere''
effect.

Secondly, given the present integrated luminosity collected at the LHC, the
second resonance is too heavy to be seen unambiguously by both experimental
collaborations and in all possible channels. In retrospect, one should
remember the discovery of the 125~GeV resonance in 2012, which was initially
seen by ATLAS and CMS predominantly in the $h\to \gamma\gamma$,
$h \to ZZ \to$ four-charged-leptons channels, and confirmed in the $h\to WW$
channel (with lower significance). However, it was not seen in the dominant
$b \bar b$ channel and in the important $\tau ^+ \tau^-$ channel, which were
expected to be quite sensitive. The channels crucial for the discovery, with
the statistics available at that time, were those in which the final states
were fully reconstructed and contained photons or $e^+ e^- , \, \mu^+ \mu^-$
pairs, providing the best invariant-mass resolution. Presumably, this
continues to be the case even in the search for a high-mass neutral resonance,
so that the absence of signals in potentially sensitive channels, but with
lower invariant-mass resolution, should not be surprising.

Thirdly, the statistical significance of deviations from the background should
be evaluated by taking into account the phenomenology of a resonance
that can produce {\it both} \/excesses and defects of events. 

With these premises, our review of LHC data is summarised next: 

\begin{itemize}
\item 
The ATLAS data for the four-lepton channel, both for the cross section
and the statistically dominant class of ggF-low events, 
show deviations from the background with a definite excess-defect
sequence which are typical for a resonance; see panels a) and b) of Fig.~2.
By subtracting from the cross-section data the non-ggF background, a fit with
Eq.~(\ref{sigmat}) gives a mass $M_H= 677^{+30}_{-14}$~GeV. The combined
statistical significance of the observed deviation is $3\sigma$. 
\item 
The ATLAS inclusive $\gamma\gamma$ events indicate a $3\sigma$
excess at $684$~GeV; see panel c) of Fig.~2. A fit to the data with
Eq.~(\ref{sigmat}) (see panel d) of Fig.~2) yields a resonance mass 
$M_H=696\,(12)$~GeV. 
\item 
For the $S/B$ in the CMS four-lepton channel (see Table~\ref{CMSleptontable}),
by considering the three values at 675, 690, and 705~GeV, one finds a
combined deviation of $2\sigma$. The fitted mass (see
Fig.~\ref{CMSvecchiofit}) comes out at $M_H= 692^{+17}_{-12}$~GeV.
\item 
The CMS-TOTEM $\gamma\gamma$ events produced in $pp$ diffractive scattering
indicate an excess of $3\sigma$ in the region of invariant mass
$M_H=650\,(40)$~GeV (see Fig.~\ref{diffractive}). 
\item 
The ATLAS data for top-quark pair production, indicate small excesses at
a mass of $675\,(75)$~GeV, which are more evident when the tracks of the
final leptons are at large angles; see Fig.~\ref{ttbar}. The statistical
significance is $1\sigma$.
\item 
ATLAS measurements in the $b \bar b +\gamma\gamma$ channel
\cite{ATLAS_BBGG_paper} to constrain the cross section
$\sigma^{\rm exp}(pp \to X \to h h)$  indicate the same excess-defect pattern
observed in the four-lepton channel by both ATLAS (see panels a) and b) 
Fig.~2) and CMS (see Fig.~\ref{CMSvecchiofit}). Since this is the
characteristic signature of background-resonance interference, here the
resonance mass would be $M_H\sim 675\,(25)$~GeV. We have also shown that the
importance of these ATLAS measurements has been overlooked. In fact, one can
construct particular combinations of the cross sections in consecutive bins
where all theoretical uncertainties practically vanish. The combined
statistical significance is thus large, 
at the level of about $3.8\sigma$, implying that the observed
deviations cannot be simple statistical fluctuations. This is
even more true as one can use the tendentially low $S/B$ ratio, past the
resonance peak and observed by both ATLAS and CMS in the four-lepton channel,
to explain the sizeable reduction of $b \bar b +\gamma\gamma$ events seen
by ATLAS in the same region of invariant mass.
\end{itemize}

Since the above determinations are all well aligned within
their respective uncertainties, we can combine the mass values and obtain
$(M_H)^{\rm comb}\sim 685\,(10)$~GeV, in very good agreement with our
prediction $(M_H)^{\rm Theor}=690\,(30)$~GeV. 
Due to the modest correlation of the above measurements, we could also
attempt a rough estimate of the combined statistical evidence 
through the sum of the squares of the individual sigmas. The combined results,
at the level of about $5.8\sigma$ from ATLAS and $3.6\sigma$ from CMS, 
definitely exclude an interpretation in terms of statistical fluctuations.

Thus, by increasing the statistics and refining the analysis, we expect the
second resonance to also show up in other channels. But these other channels,
where the second resonance has not yet been seen, cannot represent an argument
to exclude its existence. As an example, let us consider the process
$H\to WW \to 2l 2\nu$. As explained, the second resonance is essentially
produced via the ggF mechanism. Therefore, when comparing with the existing
CMS measurements \cite{CMS_X_WW_PAS}, the second resonance is in the class
of models where the VBF production mode is irrelevant. This is the case
$f_{\rm VBF}= 0$ in Fig.~4 (top left) of Ref.~\cite{CMS_X_WW_PAS}. From the
numbers reported in our Sec.~\ref{pheno}, namely a partial width
$\Gamma(H\to WW)\sim 3.27$~GeV and a total width
$\Gamma(H\to {\rm all})\sim 25\div 35$~GeV, we find a branching ratio
$B(H\to WW)\sim 0.11\,(2)$. Thus, for a ggF production cross section
of about 1~pb, we expect a resonant contribution
$\sigma(pp\to H\to WW\to 2 l 2\nu)\sim 5\,(1)\times 10^{-3}$~pb, well
consistent with the CMS 95\% upper limit of $0.02\div 0.03$~pb around 700~GeV.
On the other hand, we could also consider another CMS search for heavy
resonances $X$, viz.\ through the chain $X \to hh\to b\bar b WW$. From Fig.~18
(upper panel) of Ref.~\cite{CMS_X_bbWW}, the $S/B$ ratio is seen to decrease
from about 1.5 at 600~GeV down to about 0.5 at 750~GeV. Here, the latter
2$\sigma$ defect would be consistent with the previous average determination
$\langle S/B \rangle= 0.65\,(15)$, observed by both ATLAS and CMS in the
four-lepton channel, as well as by ATLAS in the $b \bar b +\gamma\gamma$ final
state, past the resonance peak. As such, it could be brought in support of our
picture. 

Analogous considerations could be applied to other samples of data where the
weakness of the expected signal and/or the low statistics do not allow for
stringent tests. Instead, a serious problem is that, nowadays, experiments are
compared to substantial modifications of the Standard Model (such as
explicit additional Higgs bosons, supersymmetric extensions, extra
space-time dimensions, \ldots). However, no attention is paid to the simplest
idea, namely that the same SM Higgs field may exhibit a richer pattern of
mass scales, like when SSB in $\Phi^4$ theory is described as a (weak)
first-order phase transition. In view of the sizeable deviations we have
pointed out, we hope that the experimental groups will now also consider this
other possibility.

\end{document}